\documentclass[a4paper,12pt,epsfig]{epl}\usepackage{epsfig}
\title{Electrical transport in deformed nanostrips: 
electrical signature of reversible mechanical failure}
\shorttitle{Electrical transport in deformed nanostrips}
\author{Soumendu Datta\inst{1}\thanks{E-mail:\email{soumendu@bose.res.in}}
\and Debasish Chaudhuri\inst{1}\thanks{E-mail:\email{debc@bose.res.in}}
\and Tanusri Saha-Dasgupta\inst{1} \thanks{E-mail:\email{tanusri@bose.res.in}}
\and Surajit Sengupta\inst{1}\thanks{E-mail:\email{surajit@bose.res.in}}}
\shortauthor{S. Datta {\em et. al.}}
\institute{
\inst{1}S.N. Bose National Centre for Basic Sciences - Block JD, 
Sector III, Salt Lake, Kolkata 700 098, India
}

\begin{document}

\maketitle
\begin{abstract}
We calculate the electrical conductivity of a thin crystalline strip of atoms
confined within a quasi one dimensional channel of fixed width. The conductivity
shows anomalous behavior as the strip is deformed under tensile loading. 
Beyond a critical strain, the solid fails 
by the nucleation of alternating bands of solid and {\em smectic} like 
phases accompanied by a jump in the conductivity. Since the failure of 
the strip in this system is known to be reversible, the conductivity 
anomaly may have practical use as a sensitive strain transducer.   
\end{abstract}

{\em Introduction:}
Deformation of nano meter sized wires and bars have been studied, 
using theoretical analysis as well as experiments, 
extensively in recent times\cite{nanostuff,nems,nano-wire}. Such studies 
are useful both for 
understanding deformation mechanisms in general and for their relevance 
in the construction of nano devices\cite{nems}. Single crystal nano bars and strips have
been shown to fail on tensile loading conditions by the familiar necking 
mechanism\cite{nano-wire} where an elastic instability leads to a reduction of the cross 
section of bar. While necking in bulk samples\cite{Cahn-Haasen} occurs along with extensive 
plastic deformation caused by the motion of dislocations, in nano strips and 
beams, dislocations cannot be nucleated because of much higher elastic energy 
costs\cite{nano-disloc}. This leads to novel layering transitions where the solid thins 
down layer by layer\cite{nano-wire} and finally fractures after attaining the thickness 
of an atomic chain. 
\vskip .1cm

\noindent
The situation is somewhat different if the nano sized solid is confined within
a rigid channel\cite{debc} so that the necking transition is prevented. In this case, with imposition of an external strain parallel to the confining walls,
the solid fails by a series of layer transitions where the number of crystalline
layers decreases by one, accompanied by the nucleation of bands of a fluid 
with strong orientational order. The remarkable fact is that this transition 
is completely reversible, such that a decrease of the tensile strain, 
immediately causes these failure bands to disappear and the solid heals 
itself automatically. In this Letter, we look at the electrical conductivity
of a nano solid undergoing such a transition. Our motivation is to explore 
the possibility of a strong electrical signal at the reversible transition. 
We hope that such a signal, if it exists, would be useful for designing  
nano electro-mechanical devices\cite{nems}.        

{\em The model:}
Since our aim here is to explore general principles rather than evaluate
the properties of any particular system in any great detail, we have chosen
a simple model system in two dimensions. Our calculations may be directly 
relevant for a strip of atoms adsorbed on a flat substrate and confined 
within a narrow straight channel (see Fig. 1 (a)), large enough to accommodate only a few atomic 
layers. The system geometry is generated by assuming hard disk ``atoms''  
where particles $i$ and $j$, 
interact with the {\em effective} interatomic potential 
$V_{ij} = 0$ for $|{\bf r}_{ij}| > {\rm d}$ and
$V_{ij} = \infty$ for $|{\bf r}_{ij}| \leq {\rm d}$, where ${\rm d}$ is
the hard disk diameter and ${\bf r}_{ij} = {\bf r}_j - {\bf r}_i$ the
relative position vector of the particles\cite{alder,jaster}.
In three dimensions, the corresponding hard sphere system has
been used\cite{ashcroft} in the past to model electrical properties of 
simple liquid metals with some success. 
The pure hard disk free energy is entirely entropic in
origin and the only thermodynamically relevant variable for a system of 
$N$ atoms in an area $A$ is the number density
$\rho = N/A$ or the packing fraction $\eta = (\pi/4) \rho {\rm d}^2$.
Accurate computer simulations\cite{jaster} of hard
disks show that for $\eta > \eta_f = .719$ the system exists as a triangular
lattice which melts below $\eta_m = .706$. We consider 
a narrow channel in two dimensions  of width $L_y$ defined by
hard walls at $y = 0$ and $L_y$ ($V_{\rm wall}(y) = 0$ for
$ {\rm d/2} < y < L_y - {\rm d/2}$ and $ = \infty$ otherwise) and
length $L_x$ with $L_x \gg L_y$. Periodic boundary conditions are assumed
in the $x$ direction.
\vskip .1cm

\noindent
 Once the system geometry is generated by means
of Monte Carlo simulations of ``hard disk" atoms, we use the generated
structure as the underlying atomic arrangement for which 
electrical transmittance is computed. A similar treatment has been used 
also in Ref.\cite{ashcroft} viz. using structural information from the hard 
sphere system as inputs to a calculation of electrical properties of 
liquid metals. For computation of electrical 
transmittance, a tight-binding form of the electronic Hamiltonian
\begin{displaymath}H = \sum_i\sum_jt_{ij}|i\rangle\langle j| \end{displaymath}is assumed,
with hopping interactions $t_{ij}$ between atoms $i$ and $j$. 
Two different forms of the distribution (Fig. \ref{distribution}) have been considered, 
to study the influence of
hopping strength distribution on the transmittance. The considered 
distributions have a simple power-law behavior of the form
$\frac{B}{{\rm r}^\alpha}$ where ${\rm r}$ is the distance between the atoms in the
unit of the lattice constant of the unstressed triangular lattice. 
$\alpha = 5$ for set I and $\alpha = 7$ for set II. B is taken as 2.0,
so that for ${\rm r} = 1.0$, the nearest neighbor separation in the 
unstressed lattice, the hopping interaction is set as 2.0 in some energy
unit. The distribution is also assumed to have a cut-off range, so that beyond
the second nearest neighbor distance the hopping vanishes.\\
 
\begin{figure}
\centering\epsfxsize=5in\epsfysize=2.2in\rotatebox{0}{\epsfbox{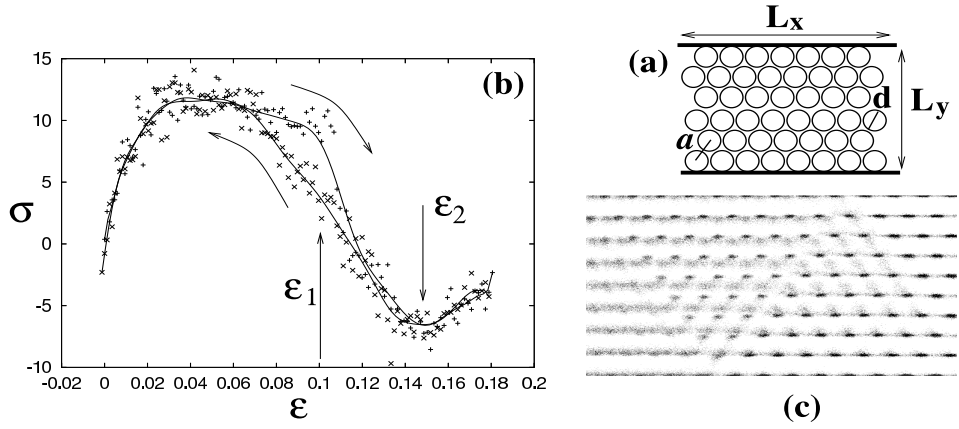}}
\caption{(a) The system geometry defining the quantities $L_x$,
$L_y$, hard disk diameter d and the lattice parameter $a$. (b)
The stress ($\sigma$) -  strain ($\epsilon$) curve as obtained from  
Monte Carlo simulations using $N = 60 \times 10$ hard disks\cite{debc}. 
The two sets of symbols
$+$ and $\times$ correspond to the stress measured while the 
strain is increased and decreased respectively. The strain values
$\epsilon_i$, $i = 1,2$ are marked. (c) Superposition of particle 
positions showing a crystal (right) - smectic interface. The number of 
layers in the crystalline region is larger by one. A smectic band 
is flanked by two such interfaces\cite{debc}. 
}\label{stress}
\end{figure}

\phantom{xx}To compute the electrical transmittance of the 
system, we attach conducting, semi-infinite, one-dimensional leads along
the horizontal direction of the sample. A set of leads are attached at
regular intervals at both ends of the sample. The purpose of these leads
is to bear the incoming, reflected and transmitted waves into and away
from the sample. The leads are described by one-dimensional, 
tight-binding, nearest-neighbor Hamiltonian of the form
\begin{displaymath}H_{lead} = V_L\sum_i(|i\rangle\langle i+1| + |i+1\rangle\langle i|)
\end{displaymath}

{\em Deformation behavior:}
\noindent
The effect of strain on the hard disk triangular solid at fixed
$L_y$ large enough to accommodate a small number of layers $n_l \sim 9 - 25$
has been studied in Ref.\cite{debc}.
The stress\cite{elast}, $\sigma = \sigma_{xx} - \sigma_{yy}$ in units of $k_B T/{\rm d}^2$,
versus strain, $\epsilon = (\eta_0 - \eta)/\eta$, curve is shown in Fig. \ref{stress}(b).
The packing fraction of the solid was taken to be $\eta_0 = 0.85$, a value deep 
in the solid phase. For $\eta = \eta_0$ ($\epsilon = 0$) the stress is
purely hydrostatic with $\sigma_{xx} = \sigma_{yy}$ as expected.
As the length, $L_x$, of the channel is increased keeping the width
$L_y$ fixed;  initially,
the stress increases linearly (Fig. \ref{stress}(b)), flattening out at the
onset of plastic behavior at $\epsilon \stackrel{<}{\sim} \epsilon_{1}$.
At $\epsilon_{1}$, with the nucleation of smectic bands,
$\,\,\sigma$ decreases and eventually becomes negative.
At $\epsilon_{2}$ the smectic phase spans the entire system and $\sigma$ is minimum.
On further increase in strain,$ \,\,\sigma$ approaches zero from below (Fig. \ref{stress}(b)) thus forming a Van der Waals loop.
If the strain is reversed by increasing $\eta$
back to $\eta_0$ the entire stress-strain curve is traced back
with no remnant stress at $\eta = \eta_0$ showing that the
plastic region is reversible. For $\epsilon_1 < \epsilon < \epsilon_2$ we observe 
that the smectic order appears within narrow bands (Fig. \ref{stress}(c)).
Inside these bands the number of layers is less
by one and the system in this range of $\epsilon$ is in a mixed phase.
The total size of such bands grows as $\epsilon$ is increased. 
\vskip .1cm

\noindent
For every value of $\epsilon$ we store a number of hard disk configurations
($\sim 1000$) which represent the instantaneous atomic positions. We use these 
configurations as structural information which are inputs to the electrical 
transport calculations to be described below. All transport quantities are averaged 
over these configurations so that our method closely corresponds to that followed in 
Ref.\cite{ashcroft}. We proceed to obtain, in this fashion, the signature of smectic 
band formation on the conductivity of the strip.  

\begin{figure}
\centering\epsfxsize=2.5in\epsfysize=2.5in\rotatebox{0}{\epsfbox{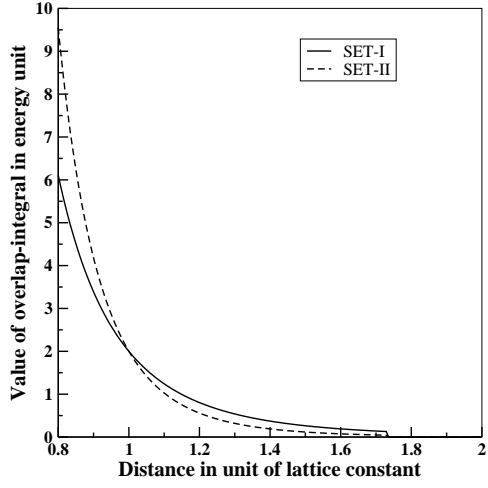}}
\caption{Two power-law distributions showing the variation of 
hopping-integral with distance}\label{distribution}\end{figure}

{\em Electrical transport:}
We compute the transmittance of the above 
described system by means of the vector recursion technique\cite{gordin1}.
The essence of the vector recursion technique is the block tridiagonalization 
of the system Hamiltonian by changing to a new orthogonal set of vector 
basis, with the restriction that the lead Hamiltonian remains unchanged. 
In this last aspect, it differs from the standard Lancos method\cite{haydock2}. The 
numerical stability of this method\cite{gordin2} has been established in studying problems 
related to Anderson localization and quantum percolation model previously\cite{indra}.
Below we describe the method briefly.\\ \phantom{xx}
A representation of the original basis is column vectors, $\{ \vert m \rangle \}$ of length
2$\cal{N}$, where $\cal{N}$ = $N$/2. Let us consider for the sake of demonstration, we have two
leads, one incoming and another outgoing connected to opposite ends of the sample at positions
$\vert 1 \rangle$ and $\vert 2\cal{N} \rangle$. A representation of the new vector basis is
then matrices of size 2$\cal{N} \times$ 2. The members of the new basis are generated in the following
way. The lead states are chosen to be,

\[
\vert \Phi_{n} \} = \left(\begin{array}{c}
\vert n \rangle \\
\vert 2{\cal N} - n + 1 \rangle \\
\end{array} \right)
\]

with $n = 0, -1, -2, \ldots, \infty$. The starting state within the system in chosen to be 

\begin{displaymath}
\vert \Phi_{1} \} = \left(\begin{array}{c}
\vert 1 \rangle \\
\vert 2\cal{N} \rangle 
\end{array} \right)
\end{displaymath}

where $\vert 1 \rangle$ and $\vert 2 \cal{N} \rangle$ are the positions where the incoming
and outgoing leads are attached. 
The subsequent members of the basis are generated from\begin{displaymath}
B_2^\dagger|\Phi_2\} = ( H - A_1)|\Phi_1\}\hskip 8.9cm
\end{displaymath}
\begin{displaymath}
B_{n+1}^\dagger|\Phi_{n+1}\} = ( H - A_n)|\Phi_n\} - B_n|\Phi_{n-1}\}
\phantom{xx} for\phantom{xx} n \ge 2\hskip 4.0cm({\bf1})
\end{displaymath}\\ \phantom{xx}
The matrix inner product is defined as \begin{displaymath}
\{\Phi\chi\}^{\mu\nu} = \sum_{i=1}^M\Phi_i^\mu\chi_i^\nu \phantom{xxx}
and\phantom{xx} orthogonality\phantom{xx} as\phantom{xx} \{\Phi|\chi\} = I
\end{displaymath}\\ \phantom{xx}
It can easily be shown that the $2\times2$ matrices $A_n$ and $B_n$ 
are block-tridiagonal members of the matrix representation of the 
Hamiltonian in the new basis:
\begin{displaymath}
A_n = \{\Phi_n|H|\Phi_n\}\hskip 0.8cm
B_n = \{\Phi_{n+1}|H|\Phi_n\}
\end{displaymath}
so that the transformed Hamiltonian matrix can be divided in 2 x 2 blocks,
with only non-zero diagonal and subdiagonal blocks.

The wavefunction $|\Psi\}$ may be represented in this new basis by a set 
$\{\psi_n\}$ so that $|\Psi\} = \sum_n\psi_n|\Phi_n\}$. These 
wavefunction amplitudes $\psi_n$ also satisfy an equation identical 
with (1).\\ \phantom{xx} The solution of the Schr\"{o}dinger equation in the leads are traveling 
Bloch waves of the form\begin{displaymath}
\sum_mAexp(\pm im\vartheta)|m\rangle \end{displaymath}
As the wave travels in the leads, the phase of its wavefunction changes by 
$\vartheta$, where \begin{displaymath}\cos\vartheta = E/2V_L 
\end{displaymath}
E being the energy of the incoming electron\cite{comment1}. In the incoming lead
there will be an incoming wave of the form $\sum exp(+ im\vartheta)|m\rangle$
and a reflected wave of the form $\sum r(E) exp(- im\vartheta)|m\rangle$.
In the output lead there will be a transmitted wave $\sum t(E) exp(- im\vartheta)|m\rangle$
\cite{comment2}, where $r(E)$ and $t(E)$ are the complex reflection and transmission
coefficients. 
The boundary conditions may then be imposed 
from the known solution in the leads: 
\begin{displaymath}
\psi_0 = \left(\begin{array}{c}
1+r(E)\\
t(E)
\end{array} \right)
\end{displaymath}
\begin{displaymath}
\psi_1 = \left(\begin{array}{c}
exp(i\vartheta)+r(E)exp(-i\vartheta)\\
t(E)exp(-i\vartheta)
\end{array} \right)
\end{displaymath}
The amplitude at the $n$th basis $\psi_n$ may be written as
\begin{displaymath}\psi_n=X_n\psi_0+Y_n\psi_1,\end{displaymath}
where $X_n$ and $Y_n$ satisfy the same recurrence relation as (1) with $EI$ replacing 
$H$ and also satisfy the boundary conditions $X_0 = I$ and $X_1 = 0$, while 
$Y_0 =0$ and $Y_1 = I$. Note that $X$ and $Y$ are $2 \times 2$ matrices.\\ \phantom{xx}
This new basis terminates after $\nu = \cal{N}$ steps, as the rank of the space spanned 
by the original tight-binding basis remains unchanged after the transformation. 
Hence the recursion also terminates after $\nu$ steps. This gives an additional boundary 
condition\begin{displaymath}X_{\nu+1}\psi_0+Y_{\nu+1}\psi_1=0_{2\times 2}\end{displaymath}
If we now interchange the incoming and outgoing leads, we get a similar pair of equations for
$r^\prime$ and $t^\prime$, the transmission and reflection coefficients for wave incident
from the second lead. Time reversal symmetry demands that $t$ must be same for waves of the same energy
incident from either lead so that $t$ = $t^\prime$. Solving these equations for the
scattering S-matrix\cite{supriyo} for the sample region one have,

\[
S = -(X_{{\cal N}+1} + Y_{{\cal N}+1}exp(-i\vartheta))^{-1}(X_{{\cal N}+1} + Y_{{\cal N}+1}exp(i\vartheta))
= \left(\begin{array}{cc}
r \enskip \enskip t \\
t  \enskip \enskip r^\prime 
\end{array} \right)
\]

Generalization of this methodology for the multi-lead case, as is the case for the present study,
with $M$ number of incoming leads and $M$ number of outgoing leads is now a trivial task.
The representation of new vector basis states formed out of repetitive application of recurrence
relation are now matrices of sizes 2${\cal N} \times$ 2M with the first member chosen as \begin{displaymath}
|\Phi_1\} = (|i_1\rangle|i_2\rangle \dots|i_M\rangle, |o_1\rangle|
o_2\rangle\dots|o_m\rangle)\end{displaymath} where $|i_k\rangle$ and $|o_k\rangle$ 
are the positions at which the 
incoming and outgoing leads attach to the system.The 2M $\times$ 2M matrices $A_n$ and $B_n$ are the block
tridiagonal representations of the Hamiltonian in the new basis. The termination of the new
basis occur after $\nu = 2\cal{N}$/2M steps with the scattering S-matrix given by,
 \begin{displaymath} S =\left(\begin{array}{ccccccc}r_{11}& r_{12} & 
\ldots & r_{1M} & t_{2M,1}^\prime & \ldots & t_{2M,M}^\prime\\\vdots & \vdots & 
\ldots & \vdots & \vdots & \ldots & \vdots\\r_{M,1} & r_{M,2} & \ldots & r_{M,M} & 
t_{M+1,1}^\prime & \ldots & t_{M+1,M}^\prime\\t_{M+1,1} & t_{M+1,2} & \ldots & 
t_{M+1,M} & r_{M,1}^\prime & \ldots & r_{M,M}^\prime\\\vdots & \vdots & \ldots & 
\vdots & \vdots & \ldots & \vdots\\t_{2M,1} & t_{2M,2} & \ldots & t_{2M,M} & 
r_{11}^\prime & \ldots & r_{1M}^\prime\end{array}\right)  \end{displaymath}
where we denote the reflection coefficient of the wavelet coming in from the $i$th 
incoming lead and reflected into the $j$th incoming lead by $r_{ij}(E)$, 
and the transmission coefficient of the same wavelet transmitted into the 
$j^\prime$ outgoing lead as $t_{ij^\prime}(E)$.\\ \phantom{xx}
The transmittance of the wavelet coming from the $i$th incoming channel is given by
\begin{displaymath}T_i(E)= \sum_{j\in O}|t_{ij}(E)|^2 \phantom{xxxxx}and\phantom{x} the\phantom{x} total\phantom{x} transmittance\phantom{x} T = \sum_{i\in I} T_i\end{displaymath}
Here $I$ and $O$ denote the sets of incoming and outgoing leads, respectively.

\begin{figure}[h]
\centering\epsfxsize=4.0in\epsfysize=2.5in\rotatebox{0}{\epsfbox{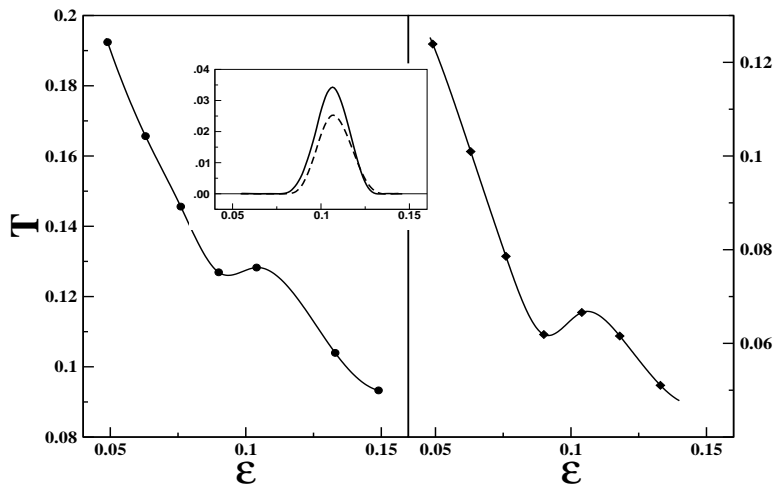}}
\caption{Variation of electrical transmittance (T) along the channel length 
with the variation of strain ($\epsilon$) for two sets of hopping integrals (left panel: set I, right panel: set II). 
Inset shows the change in transmittance
subtracting the $\sim 1/\epsilon$ behavior of the transmittance as a function of the increasing strain. The solid
and dashed lines correspond to choice of hopping integrals, set II and set I respectively.}\label{transmittance}\end{figure}

{\em Discussion and conclusions:} In Fig. \ref{transmittance}, we show the 
transmittance of the system as a function of externally imposed strain.
We notice the rather non-monotonic nature of the transmittance as the
strain is increased. With imposed strain, 
the length of the system along the horizontal direction $L_x$ increases, keeping 
the width $L_y$ fixed. This results in larger separation between atoms lying along 
the horizontal direction and therefore smaller hopping interaction giving rise to net 
reduction in the transmitted current. The transmittance going roughly as $\sim 1/\epsilon$ 
(Fig. \ref{transmittance}). This reduction continues until one reaches the 
strain value of 0.1 when the nucleation of smectic phase occurs. As explained above, 
within the smectic phase the number of atomic layers is reduced by one compared to that 
in the solid. The smectic phase with one less layer results in a decrease of the 
nearest neighbor distance {\em within} the layers and therefore to increased hopping 
interactions between atoms belonging to same layer and increased transmitted 
current along the horizontal direction as shown in the figure. 
On increasing the 
strain further, the width of the smectic band increases, thereby increasing
the atomic separation along $x$- direction and 
decreasing the hopping interaction. 
Since this change is reversible, the transmittance retraces the
curve as the strain is decreased. The change in transmittance is 
more obvious if one subtracts the overall $\sim 1/\epsilon$ behavior from the data. This
may be achieved in real devices by measuring the differential conductance between 
two similarly strained strips one of which does not undergo the layering transition. 
Since the transition depends sensitively on the width of the strip\cite{debc} this
can be easily arranged in practice. 
\vskip .1cm

\noindent
For practical applications 
one needs the change in transmittance at the nucleation of the smectic phase 
to be as sharp as possible. Our study in this context indicates that sharpness of 
the transmittance jump depends crucially on the distribution of the hopping interaction 
strength. Changing the distribution from set-I to set-II, the value of the transmittance 
decreases in general due to the reduced hopping interaction in most of the cases, but at 
the same time also leads to more pronounced jump in the transmittance at the nucleation of 
the smectic phase. This indicates the necessity of engineering of proper materials 
to exploit this phenomenon in useful devices.
\vskip .1cm

\noindent
T. S. thanks DST for Swarnajayanti fellowship. 
S.S. acknowledges financial support from DST grant SP/S2/M-20/2001. 
S.D. and D.C. thank CSIR for financial supports.


\begin{thebibliography}{0}
\bibitem{comment1}
Note that, in order to have 
propagating solutions, $|E|< 2V_L$. This sets the energy window.
The Fermi energy of the lead-sample composite system is determined
by the macroscopic lead Hamiltonian.
For our calculation we have set $E$=0, which corresponds to half filled band in an
one dimensional lead system with single s-band.
\bibitem{comment2} The process of vector recursion converts the lattice 
into an one-dimensional chain, which is then folded to clump two sites
of the chain together to define the basis set $\{ \vert \Phi_n \} \}$.
For a chain with folded configuration (see Ref. \cite{gordin1} for details)
both the reflected and transmitted waves move in opposite direction to
that of the incident wave.
\end{thebibliography}
\end{document}